# An Integrated approach for the Secure Transmission of Images based on DNA Sequences

Dr. Grasha Jacob[1], Dr. A. Murugan[2]

[1]Associate Professor, Dept. of Computer Science,
Rani Anna Govt College for Women, Tirunelveli, India
grasharanjit@gmail.com

[2]Associate Professor, Dept. of Computer Science,
Dr. Ambedkar Govt Arts College, Chennai, India

**Abstract–** *As long as human beings exist on this earth, there will be confidential images intended for limited audience. These images have to be transmitted in such a way that no unauthorized person gets knowledge of them. DNA sequences play a vital role in modern cryptography and DNA sequence based cryptography renders a helping hand for transmission of such confidential images over a public insecure channel as the intended recipient alone can decipher them. This paper outlines an integrated encryption scheme based on DNA sequences and scrambling according to magic square of doubly even order pattern. Since there is negligible correlation between the original and encrypted image this method is robust against any type of crypt attack.*

## 1. INTRODUCTION

With the growth of computer networks and the latest advances in digital technologies, a huge amount of digital data is being exchanged over various types of networks. Telemedicine has become a common method for the transmission of images and patient data across long distances. The military relies heavily on secure image transmission to provide intelligence on enemy movements and for the safety of its soldiers on the ground. In business transactions, sensible data such as pin numbers are transmitted as images. All these information are either confidential or private and ensuring security has become increasingly challenging as many communication channels are intruded by attackers. Image encryption schemes have been increasingly studied to meet the demand for real-time secure image transmission over the Internet.

Cryptography is the art encompassing the principles and methods of transforming an intelligible message into unintelligible form and then retransforming that back to its original form. The main objective of cryptography is confidentiality. Even though a cryptographic system enables in ensuring security to sensitive information, attackers and intruders have come up with various methods to crack and crash the cryptographic system. DNA sequences play a vital role in modern cryptography and DNA sequence based cryptography renders a helping hand for transmission of such confidential images over a public insecure channel as the intended recipient alone can decipher them.

Several cryptographic systems based on DNA sequences have been proposed by researchers. Qiang Zhang et. al proposed a novel image encryption algorithm based on DNA subsequence operations but with a defect of having weak ability to resist differential attack[3]. Amin et.al, introduced a DNA based implementation of YAEA. Grasha and Murugan introduced a hybrid encryption Scheme using DNA technology which had an overhead of having a key image with the same size as that of the original image and transmitting it through a secure channel [5]. This paper proposes an integrated encryption scheme that uses the DNA based implementation of YAEA and scrambling using Magic Square of doubly even order pattern to ensure double-fold security.

## 2. DEFINITIONS FOR THE PROPOSED SYSTEM

### 2.1. Cryptosystem

A cryptosystem is a five tuple (M, C, κ, ε, D) where the following conditions are satisfied.

i. M is a finite set of possible plain-text(images).
ii. C is a finite set of possible cipher-text.
iii. κ is a finite set of possible keys.
iv. For each K ∈ κ, there is an encryption rule, $E_K$ ∈ ε and a corresponding decryption rule, $D_K$ ∈ D.

Each $E_K$ : M → C and $D_K$ : C → M are functions such that $D_K(E_K(x)) = x$, for every plain-text x ∈ M.

Condition (iv) enables a user to decrypt a received cipher-text, since $D_K(E_K(x)) = x$, for every plain-text x ∈ M. For unambiguous decryption, it is required that $E_K(x_1) \neq E_K(x_2)$, if $x_1 \neq x_2$. On the other hand, if $E_K(x_1) = E_K(x_2)$, and $x_1 \neq x_2$ then decryption is not unique and hence it is impossible for a recipient to decide whether the intended plain text was $x_1$ or $x_2$ upon receipt of $E_K(x_1) = E_K(x_2)$.

### 2.2 DNA Sequence Crypt function

DNA Sequence Crypt function is a function that returns one of the many positions of the quadruple DNA sequence in the key DNA sequence file.

A one to many DNA Sequence Crypt function is a one-to-many function d(x), which has the following three properties:

i. A pointer, h maps an input quadruple nucleotide sequence, x to one of the many positions obtained in random in the key DNA sequence file.
ii. Ease of computation: Given d and an input x, d(x) is easy to compute.
iii. Resistance to guess: In order to meet the requirements of a cryptographic scheme, the property of resistance to guess is required of a crypt function with input x, $x_1$ and outputs y, $y_1$. As similar quadruple nucleotide sequence that occur in a plain text are mapped to different positions in the DNA nucleotide sequence file(one to many mapping), it is difficult for a recipient to guess the plain-text.

The sender and receiver agree on a key DNA sequence file, R which can be freely and easily downloaded from the DNA GenBank to generate the cipher text.

### 2.3. Magic Square

A magic square of order n is a square matrix or array of $n^2$ numbers such that the sum of the elements of each row and column, as well as the main diagonals, is the same number, called the magic constant (or magic sum, or line-sum), denoted by σ(M). Magic squares can be classified into three types: odd, doubly even (n divisible by four) and singly even (n is even but not divisible by four). This paper focuses only on doubly even magic square implementation and their usefulness for a cryptosystem.

## 3. INTEGRATED ENCRYPTION SCHEME

The proposed encryption scheme is an integration of a cryptosystem based on the DNA based Implementation of YAEA Encryption Algorithm proposed by Amin et. al[4] and scrambling using magic square of doubly even order pattern and to ensure double-fold security. The image to be transmitted is converted into its equivalent DNA form using the DNA digital coding technology. In this work, only images of size n x n, where n is doubly even are considered.

The reference sequence file, R is known only to the sender and the receiver and can be selected from any web-site associated with DNA sequences. The main advantage of this scheme is that the size of the encrypted image is the same as that of the original image.

In the proposed integrated encryption scheme, the image to be encrypted is first synthesized - transformed into DNA image. DNA sequences are made up of four bases – A, C, T and G. According to the Digital Coding Technology[13], C denotes 00, T – 01, A - 10 and G – 11. According to the DNA encoding method proposed for each pixel of a digital image by Qiang Zhang et.al[3], each pixel of the digital image is converted into its corresponding DNA coded value.

An intermediate image is then obtained by substituting each quadruple DNA nucleotides sequence of the translated image by one of the many positions of the quadruple nucleotides sequence which is randomly obtained from the gene sequence file. The intermediate image is scrambled according to magic square of doubly even order pattern, and the resultant encrypted image which is of the same size as that of the original image is sent to the receiver. The receiver upon receiving the encrypted image, re-scrambles it converts it into DNA image by mapping the pointers from the encrypted image onto the key DNA sequence file. The decrypted image is then obtained by Re-Synthesis.

The encryption algorithm can be summarized as follows:

**ALGORITHM DNA_MAGIC_CRYPT(X)**
*Input: X [image file] to be encrypted*
*Output: Encrypted image, Z*
1. **SYNTHESIS(X)**
2. **IF DETECT**
   Then
   **SUBSTITUTION**
   V ← DNASequenceCryptfn($P_1P_2P_3P_4$)
   Else Repeat step 2.
3. Z ← **TRANSLATION** //based on Magic Square pattern
   $P_1P_2P_3P_4 \leftrightarrow P_5P_6P_7P_8$
**End Algorithm**

The decryption algorithm can be summarized as follows:

**ALGORITHM DNA_MAGIC_DECRYPT(Z)**
*Input: Encrypted image Z*
*Output: Decrypted image, X*
1. **REV_TRANSLATION** //Reverse MAGIC square coding of Z
2. **Rev-Substitution**
   $P_1P_2P_3P_4$ ← V
3. X ← **Rev-Synthesis**
**End Algorithm**

## 4. EXPERIMENTAL RESULTS AND SECURITY ANALYSIS

To prove the trustworthiness of the proposed algorithm, experiments were performed on 200 different images of varied sizes using Matlab 2009a on DELL Inspiron ACPIx64 based notebook PC. The level of security that the proposed encryption algorithm offers is double-fold – DNA Sequence based and scrambling. The level of security that the proposed encryption algorithm offers is double-fold. Security analysis is defined as the technique of finding the weakness of a cryptographic scheme and retrieving whole or a part of the encrypted image without knowing the decryption key or the algorithm. For an encryption scheme to be good, it should be robust against statistical and brute-force attacks, and therefore the proposed method was examined for these attacks.

### 4.1 Statistical Analysis

The encrypted image should not have any statistical similarity with the original image to prevent the leakage of information. The stability of the proposed method is examined via statistical attacks - the histogram and correlation between adjacent pixels.

#### 4.1.1 Histogram Analysis

An image histogram describes how the pixels in an image are distributed by plotting the number of pixels at each intensity level. The histograms present the statistical characteristics of an image. If the histograms of the original image and encrypted image are different, then the encryption algorithm has good performance. An attacker will find it difficult to extract the pixels' statistical nature of the original image from the encrypted image and the algorithm can resist a chosen plain image or known plain image attack.

Fig 1 reveals that the histograms of the encrypted images are fairly uniform and significantly different from the original image. As the encrypted image does not provide any information regarding the distribution of gray values to the attacker, the proposed algorithm can resist any type of histogram based attacks and strengthens the security of the encrypted images significantly.

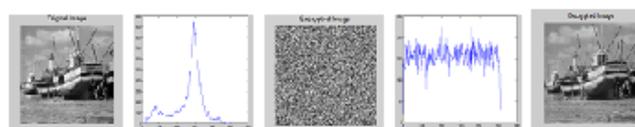

Fig.1 a) Original Image b) Histogram -original image c) Encrypted Image d) Histogram of encrypted image e) Decrypted Image

### 4.1.2 Correlation Coefficient Analysis

An efficient image encryption algorithm eliminates the correlation of pixels. Two highly uncorrelated images have approximately zero correlation coefficient.

The Pearson's Correlation Coefficient is determined using the formula:

$$r = \frac{n\sum xy - (\sum x)(\sum y)}{\sqrt{n(\sum x^2) - (\sum x)^2}\sqrt{n(\sum y^2) - (\sum y)^2}}$$

… … …    (4.1)

where x and y are the gray-scale values of two adjacent pixels in the image and N is total number of pixels selected from the image for the calculation is used to analyze the correlation.

Fig 2 (a) to (f) represent the correlation between the adjacent pixels of the original and encrypted images column-wise, row-wise and diagonal-wise and it is clear from the figures that the correlation of the encrypted images is almost uniformly distributed and gives no information to the intruder regarding the nature of the original image that is being transmitted. The two adjacent pixels in the original image are highly correlated.

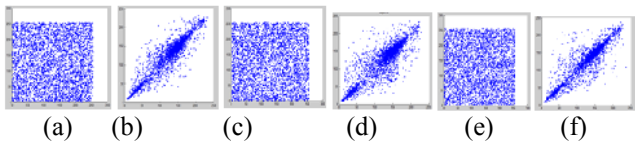

 (a)      (b)      (c)      (d)      (e)      (f)

Fig. 2 a)  Correlation coeff columnwise of  ship image
 b) Correlation coeff column-wise of Encrypted Image
 c) Correlation coeff rowwise of original image
 d) Correlation coeff  rowwise of Encrypted Image
 e) Correlation coeff diagonalwise of original image
 f) Correlation coeff diagonalwise of Encrypted Image

### 4.2   Chosen/Known-plain text attack

For encryption with a higher level of security, the security against both known-plaintext and chosen-plaintext attacks are necessary. Chosen/Known-plain text attacks are such attacks in which the intruder can access/choose a set of plain texts and observe the corresponding encrypted texts. The proposed encryption scheme was tested for chosen/known plain text attack and on execution of the algorithm the encrypted image could not be decrypted.

The algorithm for Chosen/Known–plain text attack is summarized as follows:

**Algorithm Chosen/known-plain text attack**
*Input : Plain image,C and its encrypted image C1, unknown encrypted image to be decrypted, Z1*
*Output : success/ failure*
1.   $M \leftarrow C \oplus C1$
2.   $Z = M \oplus Z1$
3.   *If Z is the correctly decrypted image of Z1*
        *then*
           *return success*
        *else*
           *return failure*
**End algorithm**

### 4.3   Guess and Determine Attack

Suppose an attacker finds the key sequence file by a guess. Even if the attacker converts the encrypted image to digital data by finding the DNA digital coding technique, the attacker will not be in a position to get the required decrypted data as the sequences were scrambled. If the attacker knows to unscramble the data, then only there is a possibility of the attacker to access the decrypted data.  However, the time complexity of this attack is quite large since there are roughly 55 million publicly available DNA sequence files.

### 4.4   Brute Force Attack

A Brute Force Attack is a strategy that can be used against any encrypted data by an attacker who is unable to take advantage of any weakness in an encryption system that would otherwise make his task easier. It involves systematically checking all possible keys until the correct key is found. The encrypted file is actually randomly generated pointers to the DNA sequence file and rarely there is a possibility of more than one quadruple nucleotide sequence pointing to the same position in the DNA sequence file. Moreover, the aspect of bio-molecular environment is more difficult to access as it is extremely difficult to recover the DNA digital code without knowing the correct coding technology used. An incorrect coding will cause biological pollution, which would lead to a corrupted image. Since there are many web-sites and roughly 55 million publicly available DNA sequences, it is virtually impossible to guess the key sequence.

### 4.5 Differential Attack

The aim of differential attack analysis is to determine the sensitivity of encryption algorithm to slight changes. If an attack is made to create a small change in the plain image to observe the results, this manipulation causes a significant change in the encrypted image and the opponent will not be able to find a meaningful relationship between the original and encrypted image with respect to diffusion and confusion. A different sequence used or a small change made to the original plain image will result in a completely different encrypted image proving that the algorithm is highly sensitive to slight changes.

### 5. CONCLUSION

. The experimental results and security analysis confirm that the proposed algorithm is easy to be implemented, can get good encryption effect, has strong sensitivity to the DNA sequence file used, and has the abilities of resisting exhaustive, statistical and differential attacks.

Integrating DNA based encryption along with magic square scrambling helps in a double fold secure transmission of confidential images. The complexity and randomness of DNA based encryption provides a great uncertainty which makes it better than other mechanism of cryptography